\documentclass[11pt]{article}
\usepackage[dvips]{graphicx,rotating}
\usepackage{epsfig,subfigure}
\usepackage{amsmath}	

\input{pubboard/babarsym}
\def\Dp      {\ensuremath{D^+}}
\def\Dm      {\ensuremath{D^-}}
\def\Dstar   {\ensuremath{D^{*}}}
\def\Dstarp  {\ensuremath{D^{*+}}}
\def\Dstarm  {\ensuremath{D^{*-}}}

\newcommand{\etal}{{\it et al.}}

 \def\er #1 #2 { $#1 \pm #2$ }
 \def\bra #1 #2 #3 #4 { $#1 ^{+#2} _{-#3} \pm #4 $ }

 \newcommand {\mhad}{\mbox{$M_{Had}$}}
 \newcommand {\cost}{\mbox{$\cos\theta_{T^*}$}}

 \newcommand {\MC}{Monte Carlo}

\newcommand{\BABARPubYear}    {02}

\newcommand{\BABARConfNumber} {25}
\newcommand{\SLACPubNumber} {9308}

\long\def\inst#1{\par\nobreak\kern 4pt\nobreak
    {\it #1}\par\vskip 10pt plus 3pt minus 3pt}

\begin{document}
\pagestyle{empty}

\begin{flushright}
\babar-CONF-\BABARPubYear/\BABARConfNumber \\
SLAC-PUB-\SLACPubNumber \\
July 2002 \\
\end{flushright}

\par\vskip 3cm

\begin{center}
{
\Large \bf
{\boldmath$ b\to s\gamma$} using a Sum of Exclusive Modes
}
\end{center}
\bigskip

\begin{center}
\large The \babar\ Collaboration\\
\mbox{ }\\
July 24, 2002
\end{center}
\bigskip \bigskip

\begin{center}
\large \bf Abstract
\end{center}
This paper describes preliminary results on the inclusive process
$b\to s\gamma$ obtained from 20.7~\invfb\ of data recorded with the 
\babar\ detector during 1999-2000. 
Signal event yields are found from a combination of twelve exclusive 
decay channels after subtracting continuum and \BB\ backgrounds. 
Cross--feed from incorrectly reconstructed $b\to s\gamma$ events is also removed. 
Branching fractions in bins of hadronic mass
are calculated using corrected Monte Carlo signal efficiencies; 
this is equivalent to measuring the gamma energy spectrum. 
We measure the first moment of the gamma energy spectrum constraining 
the HQET parameter ${\rm \overline{\Lambda}} = 0.37 \pm 0.09~(stat) \pm 0.07~(syst) \pm 0.10~(model)$~\gevcc.
A fit to the hadronic mass spectrum gives 
${\cal B}(b\to s\gamma)=(4.3\pm 0.5~(stat)\pm 0.8~(syst)\pm 1.3~(model))\cdot 10^{-4}$
for the inclusive branching fraction.  
We also constrain the HQET parameter $\lambda_1$. 
\vfill
\begin{center}
Contributed to the 31$^{st}$ International Conference on High Energy Physics,\\
7/24---7/31/2002, Amsterdam, The Netherlands
\end{center}

\vspace{1.0cm}
\begin{center}
{\em Stanford Linear Accelerator Center, Stanford University,
Stanford, CA 94309} \\ \vspace{0.1cm}\hrule\vspace{0.1cm}
Work supported in part by Department of Energy contract DE-AC03-76SF00515.
\end{center}

\newpage
\pagestyle{plain}

\begin{center}
\small

The \babar\ Collaboration,
\bigskip

B.~Aubert,
D.~Boutigny,
J.-M.~Gaillard,
A.~Hicheur,
Y.~Karyotakis,
J.~P.~Lees,
P.~Robbe,
V.~Tisserand,
A.~Zghiche
\inst{Laboratoire de Physique des Particules, F-74941 Annecy-le-Vieux, France }
A.~Palano,
A.~Pompili
\inst{Universit\`a di Bari, Dipartimento di Fisica and INFN, I-70126 Bari, Italy }
% ...

\end{center}\newpage
 \setcounter{footnote}{0}

\section{Introduction}
\label{sec:generator}

\noindent The $b \rightarrow s \gamma$ transition proceeds by an effective 
flavor-changing neutral current. The Standard Model prediction,   
in which the $t$ penguin loop gives the largest contribution,  
has been calculated to next--to--leading order~\cite{Misiak}.
Measurements of the inclusive branching fraction have been used to constrain 
new physics contributions to the decay amplitude~\cite{Hewett}.  

Since $b\to s\gamma$ is a two--body decay process, the photon energy, $E_{\gamma}$, in the \B\ rest frame, is  
related to the recoil hadronic mass, \mhad, by:

\begin{equation}
E_{\gamma} = {{M_B^2 - \mhad ^2}\over{2M_B}}.
\end{equation} 
In our analysis we fully reconstruct twelve exclusive $b\to s\gamma$ decays, and use the hadronic mass
spectrum to measure the photon energy spectrum in the \B\ rest frame.
Note that in a fully inclusive analysis~\cite{incl}, where only the photon is measured,
the gamma energy is smeared by the resolution of the calorimeter and by 
the motion of the \B\ meson in the \FourS\ rest frame. 
With our semi-exclusive approach, the gamma energy resolution 
depends only on the \mhad\ resolution, which is a few~\mevcc.

We fit the \mhad\ spectrum with the model proposed by Kagan and Neubert~\cite{KaganNeubert},
which predicts the shape of the gamma energy spectrum using heavy quark effective 
theory (HQET). 
There are two main parameters, an effective $b$ quark mass, $m_b$, 
and $\lambda_1$, which is related to the 
kinetic energy of the $b$ quark in the \B\ meson. 
The model introduces the known $K^*\gamma$ contribution~\cite{Kstar} using local 
quark-hadron duality to convert the portion of the spectrum below a cutoff 
mass into a $K^*(892)$ Breit-Wigner shape. The cutoff mass is a free parameter 
in the model, but is expected to be between 1.0 and 1.2~\gevcc. 
Above this cutoff the inclusive model does not explicitly include any higher resonances.

We find a correlated band of allowed values for the parameters $m_b$
and $\lambda_1$, and a cutoff mass of about 1.11~\gevcc. 
By measuring the first moment of the
photon energy spectrum we constrain the HQET parameter
$\rm \bar{\Lambda}$ and hence $m_b$. 
A fit to the \mhad\ spectrum with this constraint
applied to $m_b$ gives the inclusive branching fraction
and a range of values for $\lambda_1$.
There is currently much interest in the HQET parameters, 
since they are needed to extract $V_{ub}$ and $V_{cb}$ from 
semileptonic \B\ decays~\cite{Neubert}.

\section{The \babar\ detector and dataset}
\label{sec:babar}

This paper describes preliminary results obtained from 20.7~\invfb\ of data recorded with the 
\babar\ detector at the \pep2\ asymmetric $e^+e^-$ storage ring during 1999-2000. 
The data correspond to a total of $(22.74\pm 0.36) \cdot 10^6$ \BB\ pairs 
collected on the $\Upsilon(4S)$ resonance. 

The \babar\ detector is described in detail elsewhere~\cite{ref:babar}.
We briefly summarize the detector systems most relevant to the current paper.
The \babar\ detector contains a five--layer silicon vertex tracker (SVT)
and a forty--layer drift chamber (DCH) situated in a 1.5\,T solenoidal
magnetic field.
These devices detect charged particles and measure their momentum
and ionization energy loss ($dE/dx$).
The transverse momentum resolution is
$\sigma_{p_t} / p_t = (0.13 \pm 0.01) \% \cdot p_t+ (0.45 \pm 0.03)\%$,
with $p_t$ in \gevc.
Photons are detected in a CsI(Tl) crystal electromagnetic calorimeter (EMC).
The EMC detects photons with energies as low as 20~\mev.
The nominal EMC resolution for photons and electrons is
$\sigma_E / E = 2.3\% / E^{1/4} \oplus 1.9\%$, with $E$ in \gev.
The charged particle identification (PID) combines $dE/dx$ measurements
in the SVT and DCH with particle velocity measurements,
obtained with an
internally reflecting ring-imaging Cherenkov detector (DIRC) of quartz bars
surrounding the DCH.

The interactions of particles traversing the detector are simulated 
using GEANT4~\cite{geant4}. Variations in detector conditions and
beam-induced backgrounds are taken into account in these simulations.
Signal and generic background samples are used to study the effect of
the event selection criteria and to estimate the backgrounds.
The generic background simulations consist of
$e^+e^- \rightarrow q\overline{q}$ $(q=u,d,s,c)$ and \BB\ events, 
where the $b\to s\gamma$ signal events have been removed.

\section{Event Selection} 
\label{sec:selintro}

We reconstruct 12 final states\footnote{Charge conjugate modes are assumed throughout this paper.} 
formed from a high--energy photon, a charged or neutral kaon, and 1-3 pions, one of which can be neutral:
\begin{center}
$K^+\pi^-$, $K_S\pi^0$, $K^+\pi^-\pi^0$, $K_S\pi^+\pi^-$, $K^+\pi^-\pi^+\pi^-$, $K_S\pi^+\pi^-\pi^0$
\end{center}
\begin{center}
$K^+\pi^0$, $K_S\pi^+$, $K^+\pi^-\pi^+$, $K_S\pi^+\pi^0$, $K^+\pi^-\pi^+\pi^0$, $K_S\pi^+\pi^-\pi^+$
\end{center}
These make up about 50\%\ of the total $b\to s\gamma$ rate for \mhad\ $<$ 2.4~\gevcc.

We select events with at least one photon with an energy  
$1.5 < E^*_{\gamma} < 3.5~\gev$ in the $e^+e^-$ centre-of-mass (CM) frame. 
If the photon can be combined with another photon with 
energy $>$ 50~\mev\ (250~\mev) to make a $\pi^0$ ($\eta$) with a mass between 115~\mevcc\ (508~\mevcc) and
155\mevcc (588~\mevcc), it is vetoed.

In the hadronic system, a $\pi^0$ is made from two photons, with a minimum photon energy
of 30~\mev, and a minimum $\pi^0$ energy of 200~\mev. Well reconstructed tracks are used 
for charged pions, but charged kaons must satisfy additional kaon identification
criteria. Candidates for $K_S\to\pi^+\pi^-$ decays are selected 
by requiring a mass between $489$ and $507~\mevcc$, and a decay length $>2$~mm. 
These kaon selections reduce any $b\to d\gamma$ background contributions to 
a negligible level. 

The mass of the hadronic system is restricted to:
\begin{displaymath}
0.6< \mhad < 2.4~\gevcc
\end{displaymath} 
where the upper limit corresponds to a requirement on $E_{\gamma}>2.094$~\gev.
Most of the background is accounted for by events in which the photon comes from 
initial state radiation (ISR) or from $\pi^0$/$\eta$ decays that survive the veto.
We suppress these by requiring $\cost <0.7$, where \cost\ 
is the angle in the CM system between the high--energy photon candidate 
and the thrust axis of the rest of the event calculated after removing the \B\ candidate.
Further suppression is achieved by combining several event shape variables in a 
Fisher discriminant. The variables include the ratio of the 
second and zeroth Fox Wolfram moments~\cite{foxwolfram} computed in the CM and in the photon recoil system, 
the CM direction of the \B\ candidate, and 
energy flow in a 20\degrees\ cone along the photon direction 
and in a 40\degrees\ cone opposite the photon.

The hadronic system is combined with a high--energy photon, 
and accepted as a \B\ candidate if the energy in the CM system, $E^*_{B}$,
is within 150~\mev\ of the CM beam energy, $E_{beam}^*$, : 
\begin{equation}
|\Delta E| = |E^*_{B}-E_{beam}^*|<150~\mev 
\end{equation}
and the beam-energy substituted mass \mes: 
\begin{equation}
\mes = \sqrt{E_{beam}^{*2} - p_B^{*2}} > 5.21~\gevcc
\end{equation}
where $p_B^*$ is the \B\ momentum in the CM frame. 

If there is more than one \B\ candidate left in an event,
we choose the candidate with the minimum $|\Delta E|$.
Then, we rescale the measured photon energy to give 
$\Delta E = 0~\gev$, and recalculate \mes. This corrects for energy leakage and 
improves the \mes\ resolution, particularly for candidates 
without a $\pi^0$. 

\section{Background Subtraction}
\label{sec:bkgd}

The background is mainly from continuum, although \BB\ backgrounds become
significant at high \mhad\ values. 
The continuum background \mes\ distribution can be fit by an ARGUS~\cite{argus}
function. When divided into bins in \mhad, the generic continuum 
Monte Carlo samples show a slight dependence of the shape parameter of the ARGUS 
function on \mhad, which we take into account in our fits. 

The \BB\ background is mainly composed of final states with high multiplicity
and high hadronic mass, where the photon is the daughter of a $\pi^0$ or $\eta$
that decayed asymmetrically. We fit the generic \BB\ Monte Carlo samples
with an ARGUS function for the non-peaking background and a Crystal Ball 
function~\cite{CB} for the peaking background, 
such as $B\to D^{(*)}\rho^-$ where only a low--energy photon is missing.
The shape parameters of the ARGUS functions for the continuum and non-peaking 
\BB\ backgrounds are sufficiently 
similar that it is possible to use a combined ARGUS function to describe
the sum of the two backgrounds.  

\begin{figure}[htb]
  \begin{center}
   \mbox{\includegraphics[width=4.2in]{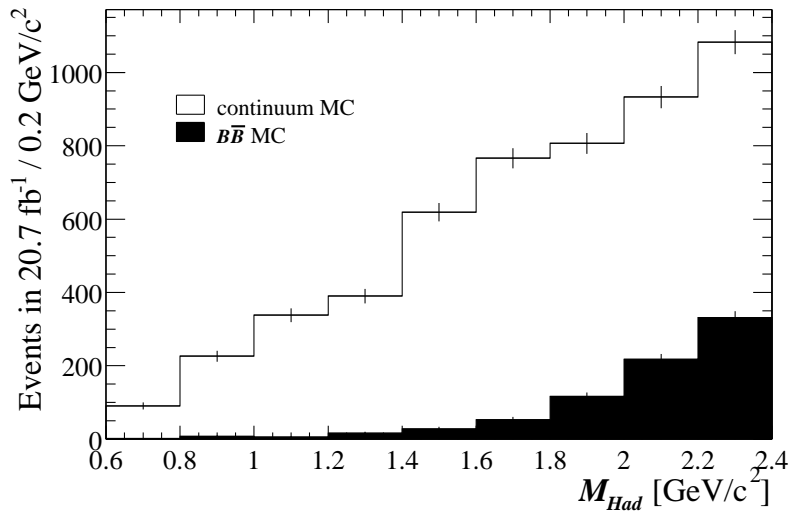}} \\
  \end{center}
\vspace{-0.8cm}
\caption{
Expected continuum and \BB\ background events in 20.7~\invfb\ from \MC\ (MC) simulation
as a function of the hadronic mass.
\label{fig:bkgmhad}}
\end{figure}

Figure~\ref{fig:bkgmhad} shows the continuum and non-peaking \BB\ backgrounds 
for $\mes > 5.21~\gevcc$, in bins of hadronic mass. The level of \BB\ background is only significant  
for $\mhad > 1.8$~\gevcc.

We treat $b\to s\gamma$ events as ``cross-feed'' background 
when they are reconstructed with the correct photon, 
but with the wrong hadronic system, since this gives an incorrect measurement of \mhad.
Cross-feed events come from three sources, which are, in decreasing order of importance:
\begin{itemize}
	\item Final states that are not considered in the analysis, e.g. $K_L$,
              $\geq 2\pi^0$, $\geq$ 5--body hadronic final states, 
              and events with $\mhad > 2.4~\gevcc$;
        \item Final states which are considered, but in which one of the final state
              particles is not detected; and
        \item Multiple candidate events where the wrong candidate is chosen. 
\end{itemize}
The cross-feed \mes\ distribution is fit with an ARGUS function plus the  
Crystal Ball function to allow for a possible peaking contribution. 
The cross-feed ARGUS function rises towards the signal region.
Note that the level of cross-feed background is proportional to the inclusive branching fraction.  
The peaking component of the cross-feed background contributes just a few events.

\begin{figure}[hptb]
  \begin{tabular}{cc}
   \mbox{\includegraphics[width=3.0in]{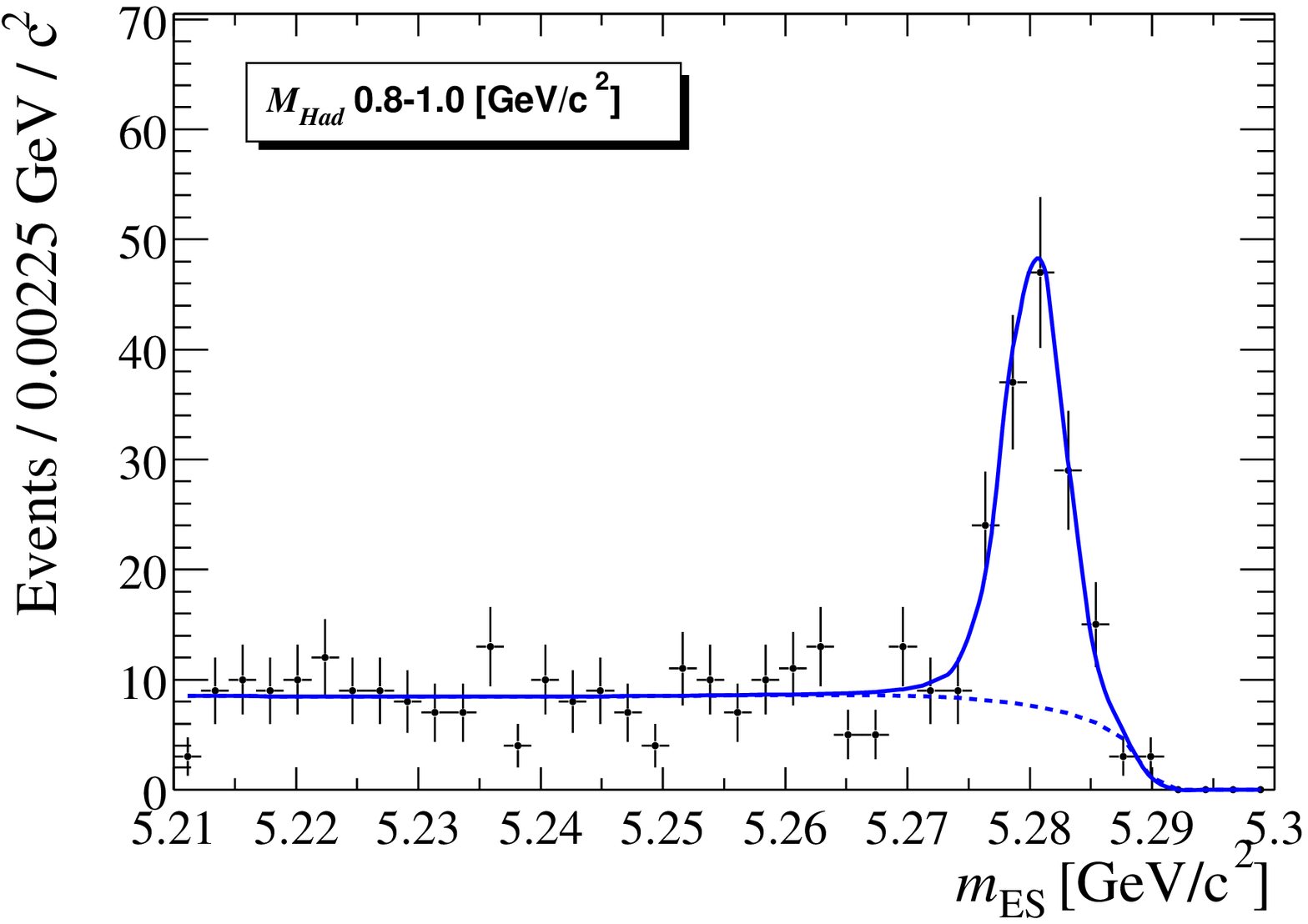}} &
   \mbox{\includegraphics[width=3.0in]{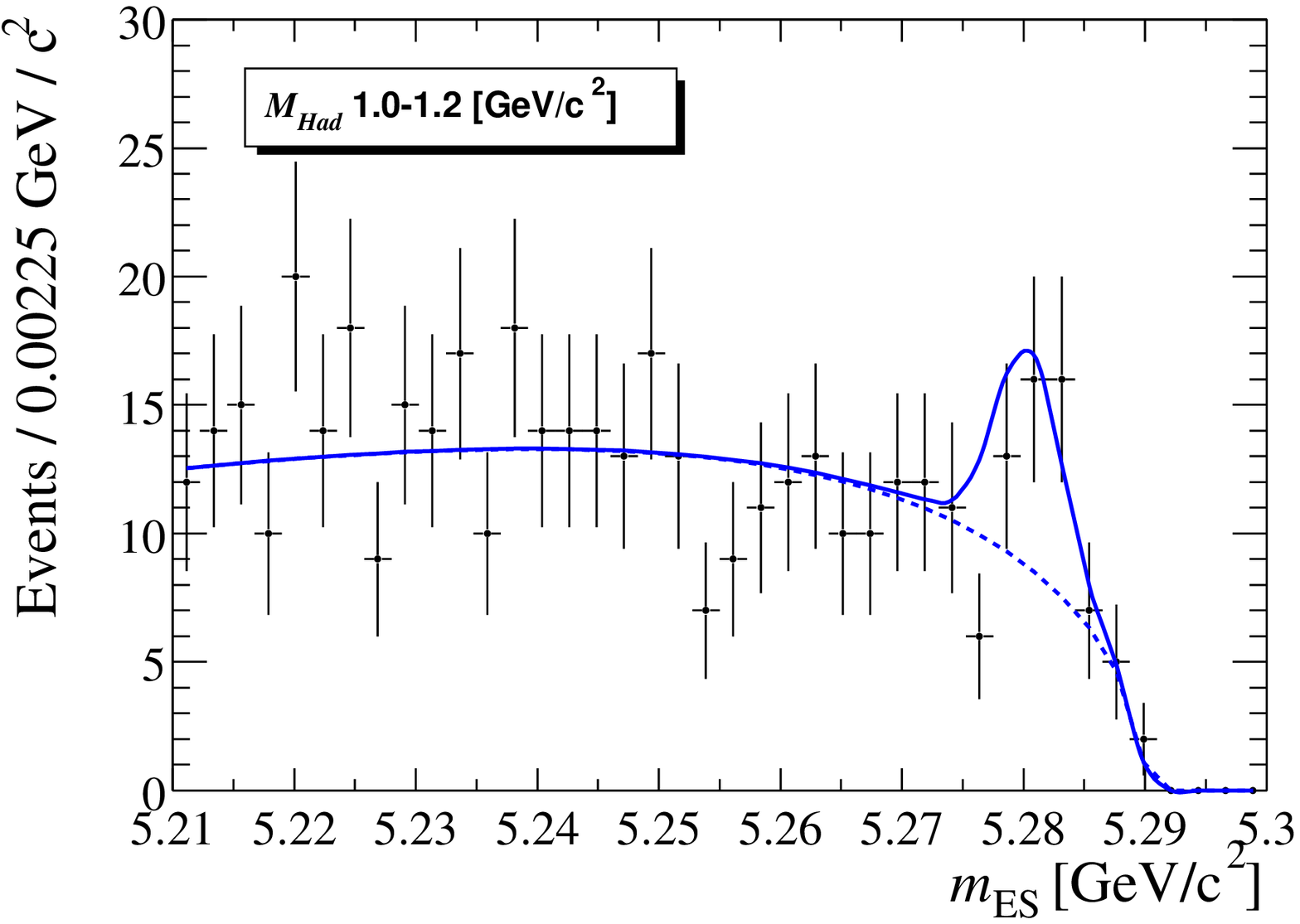}} \\
   \mbox{\includegraphics[width=3.0in]{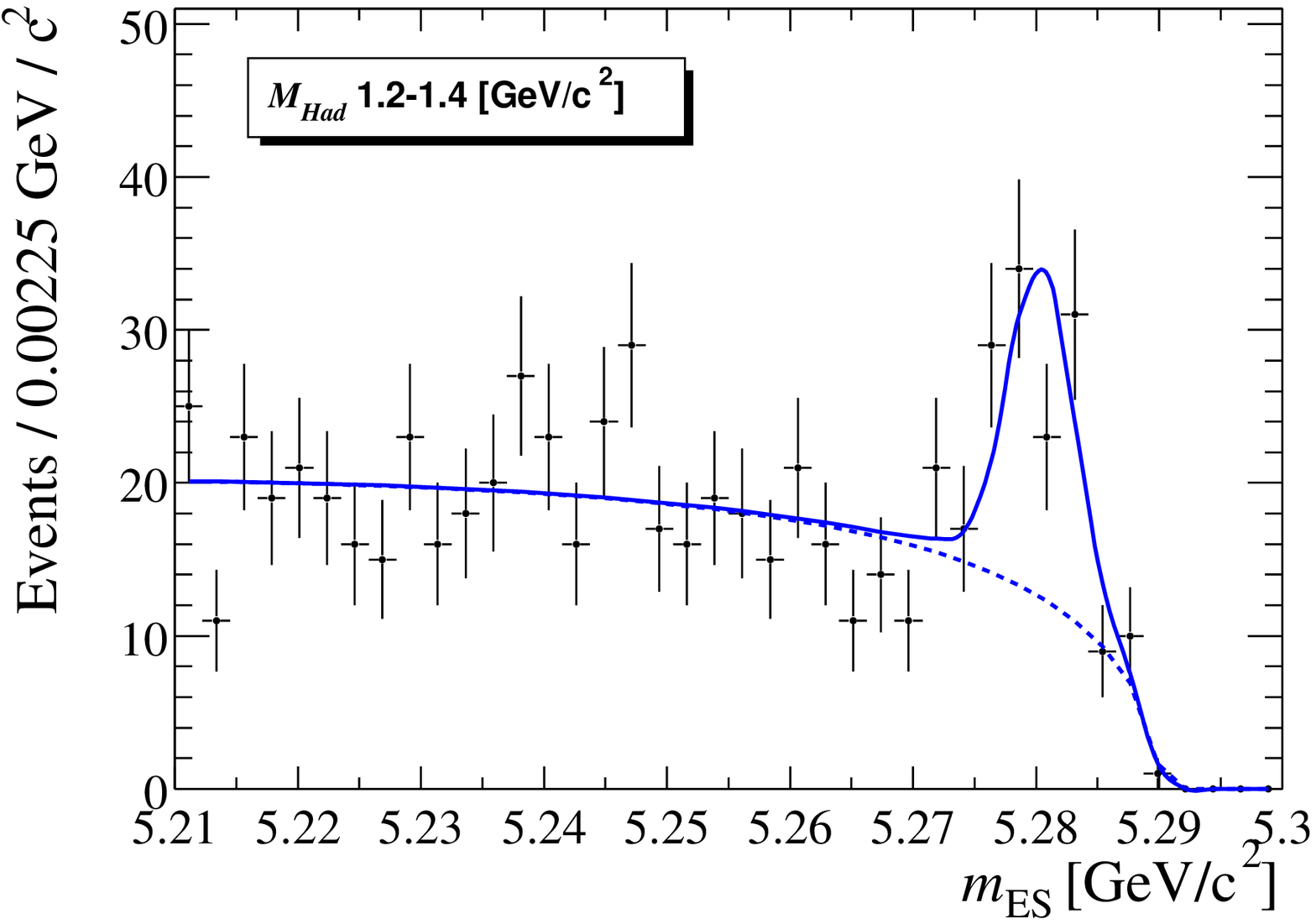}} &
   \mbox{\includegraphics[width=3.0in]{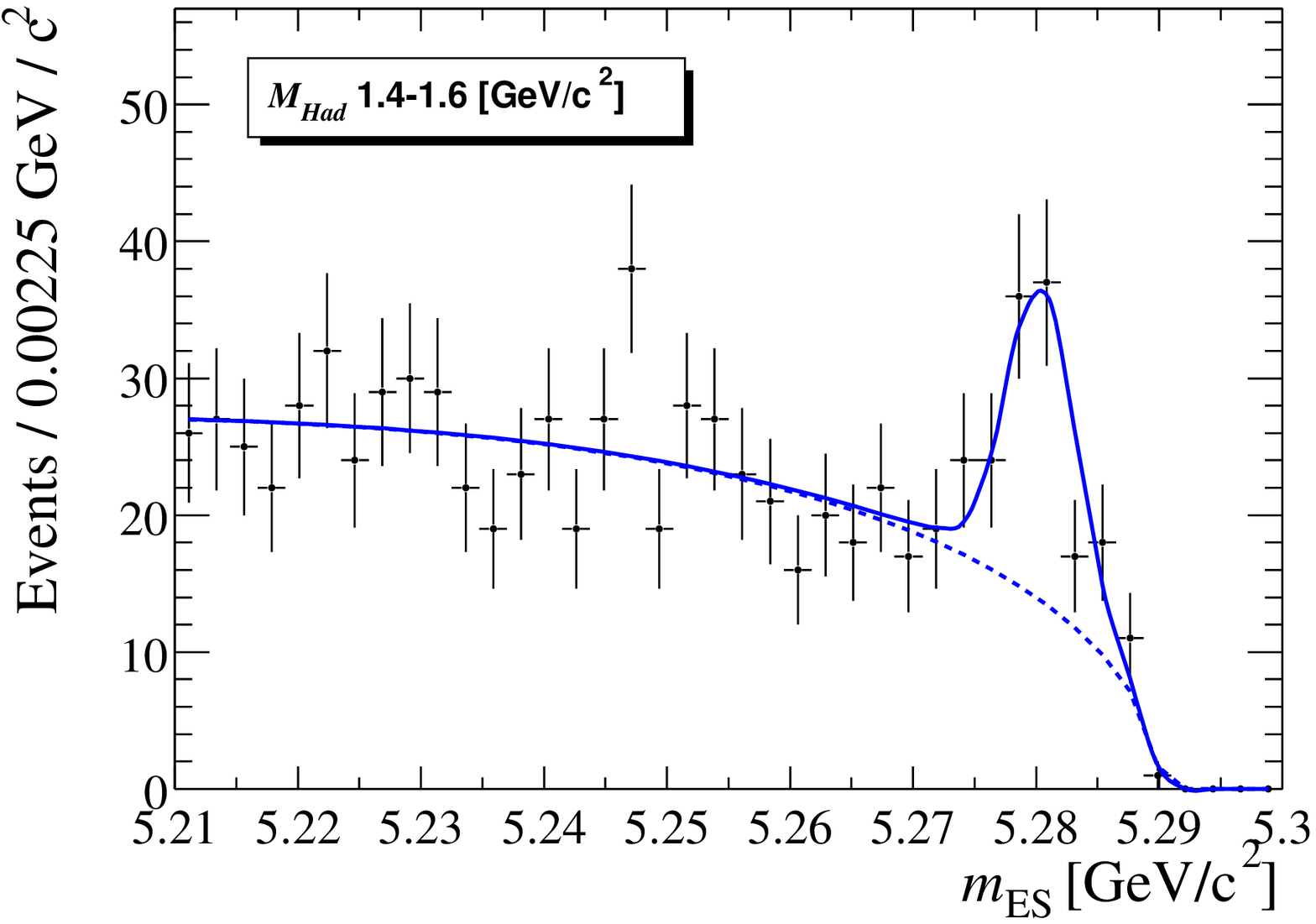}} \\
   \mbox{\includegraphics[width=3.0in]{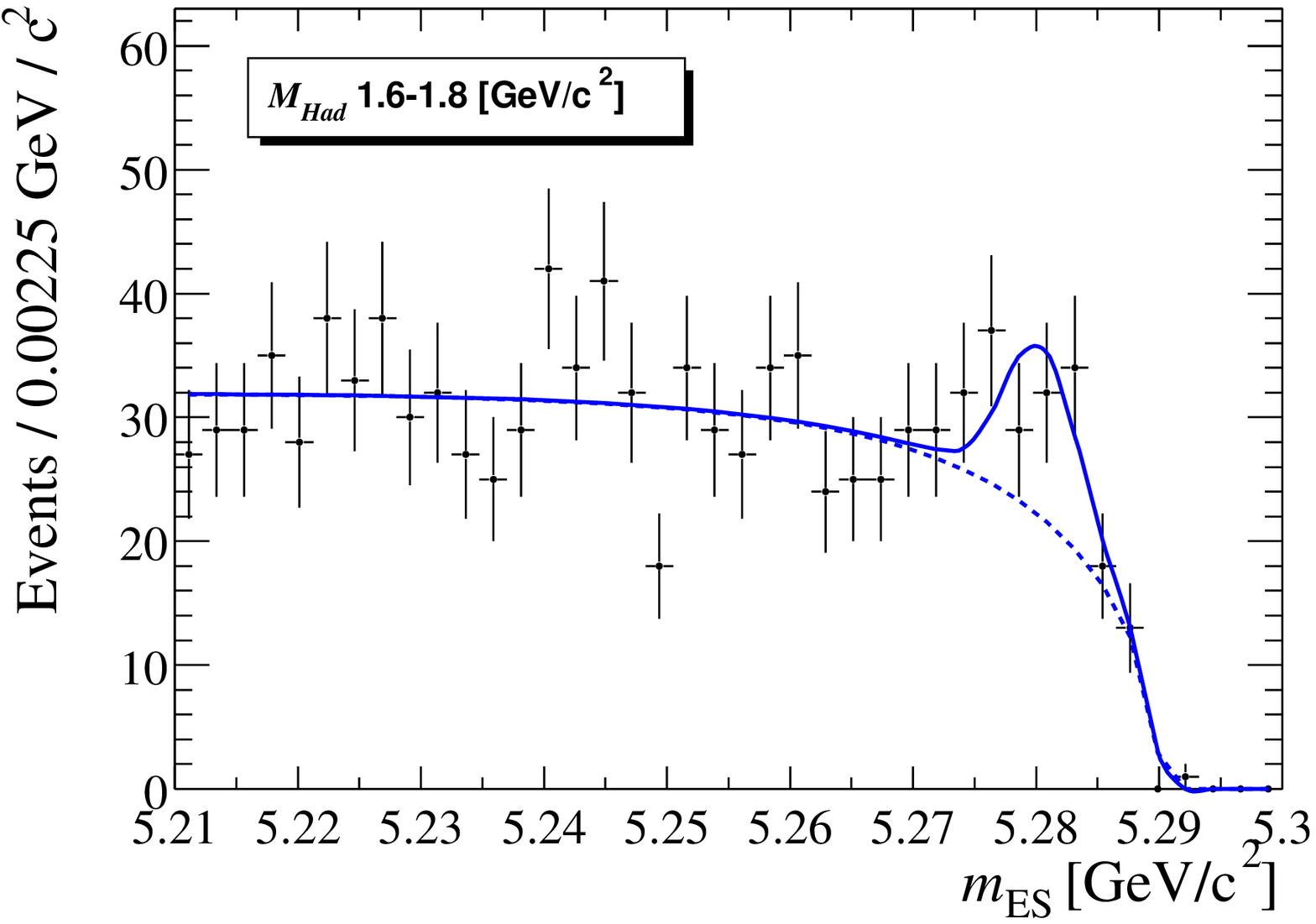}} &
   \mbox{\includegraphics[width=3.0in]{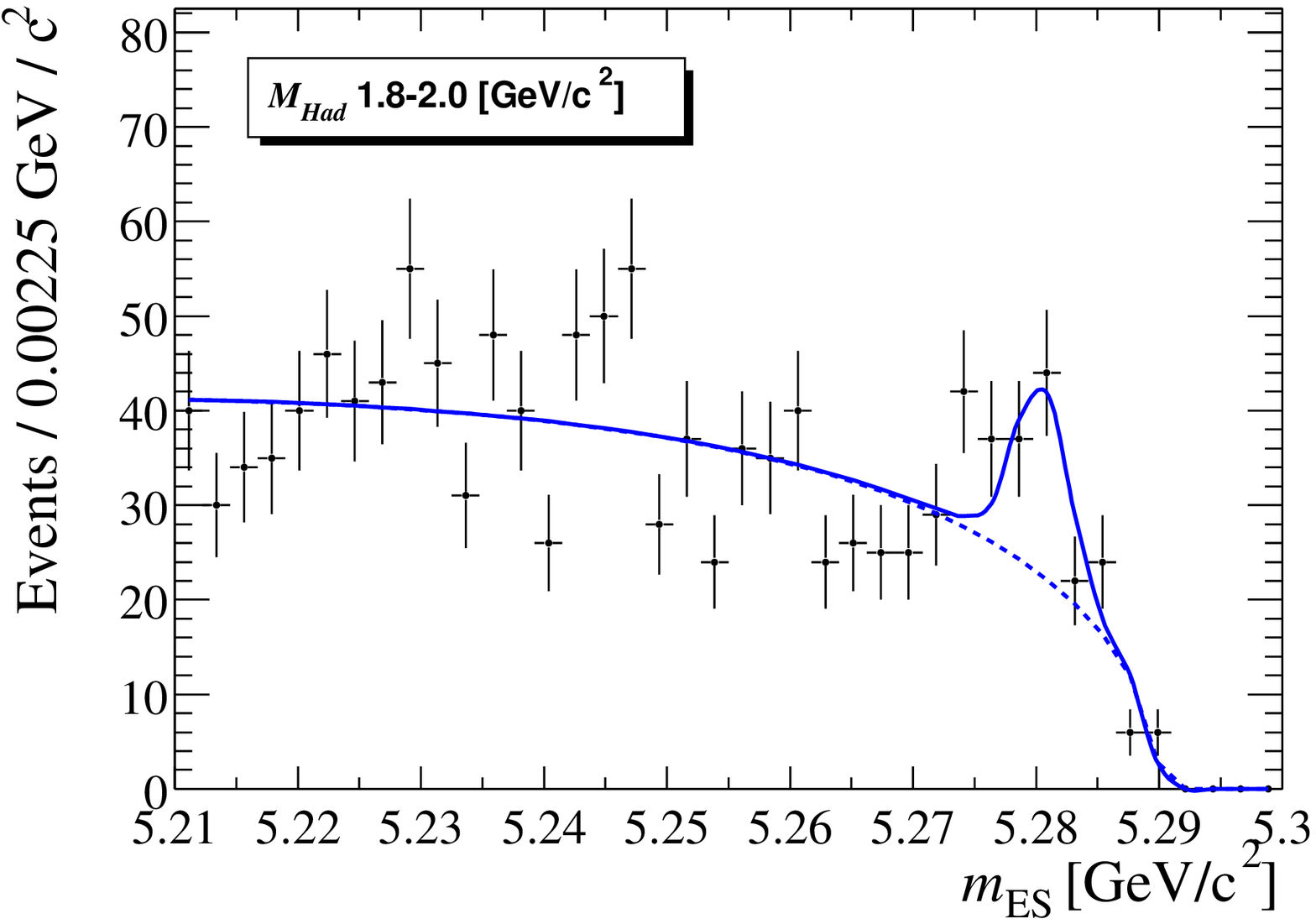}} \\
   \mbox{\includegraphics[width=3.0in]{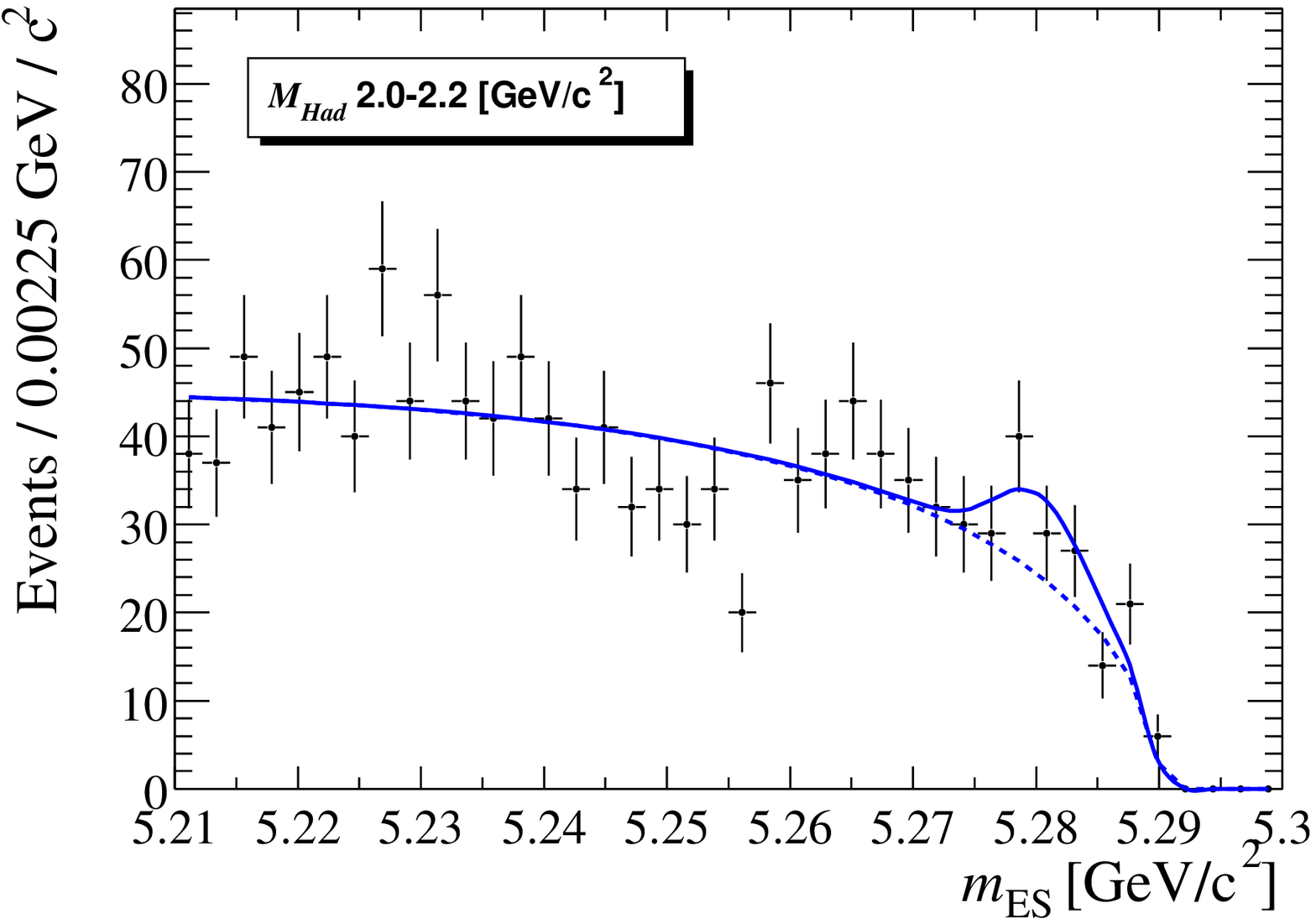}} &
   \mbox{\includegraphics[width=3.0in]{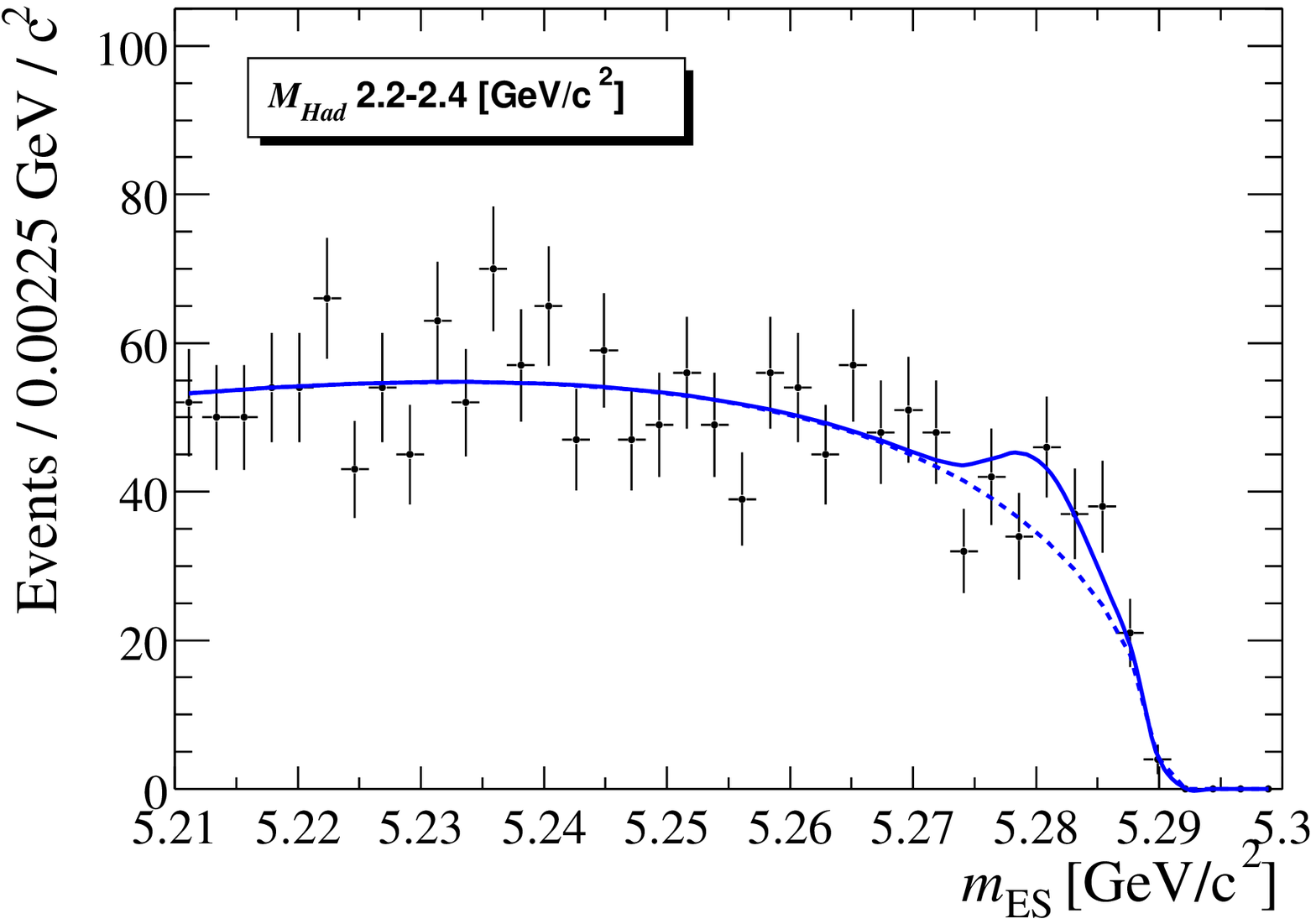}} \\   
  \end{tabular}
\caption{The fits to the data in bins of \mhad\ between 0.8 and 2.4~\gevcc.
\label{fig:fits}}
\end{figure}

Fits to the \mes\ distributions of data events are performed in each 
of the nine \mhad\ bins using a Crystal Ball function for the contributions 
that peak in the signal region and two ARGUS functions for the non-peaking 
backgrounds.
The plots of the fits to the eight \mhad\ bins from 0.8 to 2.4~\gevcc\ can
be seen in Figure~\ref{fig:fits}.

Table~\ref{tab:allyield} lists the peaking yields for the sum of all final states 
in data, \BB\ \MC\ and cross--feed, normalized to our measured inclusive 
branching fraction.

\begin{table}[htbp]
\begin{center}
\renewcommand{\arraystretch}{1.15}
\begin{tabular}{|c|rcl|rcl|rcl|} \hline
\mhad~[\gevcc]    &  \multicolumn{3}{|c|}{data } &  \multicolumn{3}{|c|}{\BB\ }        &  \multicolumn{3}{|c|}{  cross-feed  }\\ \hline
0.6-0.8 &     10 & $\pm$ &    6  & &       --  &     &    0.0 & $\pm$ & 0.1       \\
0.8-1.0 &    120 & $\pm$ &   13  & &       -- &      &    4.0 & $\pm$ & 0.5    \\
1.0-1.2 &     27 & $\pm$ &    8  & &       -- &      &    2.8 & $\pm$ & 1.0        \\
1.2-1.4 &     70 & $\pm$ &   15  & &       -- &      &    9.2 & $\pm$ & 3.2     \\
1.4-1.6 &     77 & $\pm$ &   14  & &       -- &      &    7.2 & $\pm$ & 4.4      \\
1.6-1.8 &     49 & $\pm$ &   16  & &       -- &      &    5.0  &$\pm$ & 2.9      \\
1.8-2.0 &     53 & $\pm$ &   14  &    12 & $\pm$ &   7  &    7.3 & $\pm$ & 4.1     \\
2.0-2.2 &     39 & $\pm$ &   17  &    16 & $\pm$ &  12  &   -2.0 & $\pm$ & 6.2       \\
2.2-2.4 &     41 & $\pm$ &   16  &    30 & $\pm$ &  15  &    6.7  &$\pm$ & 2.2   \\
\hline
\end{tabular}
\end{center}
\caption{ 
The data, \BB\ and cross--feed yields for all final states in the nine \mhad\ bins with their statistical errors.
\label{tab:allyield}}
\end{table}

\section{Signal Efficiency}
\label{sec:efficiency}

The signal efficiency is determined from Monte Carlo samples
using the yield for true signal events after cross-feed has been removed. 
In the first two \mhad\ bins, 0.6-1.0~\gevcc, the signal is modeled by the four exclusive 
$K^* \gamma$ modes. In the range 1.0--2.4~\gevcc, we use the inclusive $X_s \gamma$ 
model of Kagan and Neubert with $m_b=4.65$~\gevcc. 
This inclusive model uses non-resonant JETSET~\cite{jetset} 
fragmentation of the $X_S$ to produce the hadronic final states.
Changing the modeling of the 1.0-1.2~\gevcc bin to $K^*\gamma$, and varying 
the parameters of the inclusive model form part of our systematic studies. 

The overall efficiency in each \mhad\ bin is an average over the individual final states, 
weighted by the fractional contribution of that final state to the total in that \mhad\ bin.
These efficiencies have to be corrected by $\approx 10$\% for the differences between 
data and Monte Carlo detection efficiencies for $\pi^+$, $K^+$, $\pi^0$, $\gamma$ and $K_S$.
There are significant differences between the efficiencies for the individual final states
and a strong dependence of the overall efficiency on \mhad.  
To illustrate this, the efficiencies can be understood 
in terms of a 1/\mhad\ dependence of the efficiency for each final state, 
a 60\% efficiency ratio between $\pi^0$ and $\pi^+$ final states,
a 55\% efficiency ratio between $K_S$ and $K^+$ final states, and a
60\% efficiency factor between 3 and 2--body, and between 4 and 3--body hadronic systems.  

To check the fractions of each final state in the non-resonant Monte Carlo, 
we divide the sample above $\mhad =1.0~\gevcc$ into three sets of final state categories:
with and without a $\pi^0$, with and without a $K_S$, and 2, 3, 
or 4--body hadronic systems.
A check of the $K_S/K^+$ samples shows that the data and Monte Carlo values agree,
and the ratio $K_S/K^+$ is consistent with 0.5 as expected by isospin arguments.
However, we find that the fraction of $\pi^0$ final states needs to be increased 
by a factor 1.5 and the fraction of 2--body final states decreased
by a factor 0.4 in order to obtain a reasonable agreement between non-resonant
\MC\ simulation and data.
When these adjustments are applied to the fractions of final states generated by JETSET, 
they lead to a correction of $\approx 15$\% in the overall efficiency. 
Table~\ref{tab:effsig} presents the signal efficiencies after all corrections have been applied. 

\begin{table}[htbp]
\begin{center}
\begin{tabular}{|c|rcl|} \hline
\mhad~[\gevcc] &  \multicolumn{3}{|c|}{Efficiency }   \\ \hline
0.6-0.8  &  0.133 & $\pm$ & 0.009  \\
0.8-1.0  &  0.094 & $\pm$ & 0.002 \\
1.0-1.2  &  0.058 & $\pm$ & 0.004  \\
1.2-1.4  &  0.045 & $\pm$ & 0.003 \\
1.4-1.6  &  0.040 & $\pm$ & 0.002 \\
1.6-1.8  &  0.034 & $\pm$ & 0.002  \\
1.8-2.0  &  0.022 & $\pm$ & 0.002  \\
2.0-2.2  &  0.022 & $\pm$ & 0.003  \\
2.2-2.4  &  0.015 & $\pm$ & 0.003 \\
\hline
\end{tabular}
\end{center}
\caption{ The corrected signal efficiencies in bins of \mhad\ with their statistical errors.
\label{tab:effsig}}
\end{table}

\section{Branching Fractions}
\label{sec:br}

\begin{figure}[htb]
  \begin{tabular}{cc}
   \mbox{\includegraphics[width=3.2in]{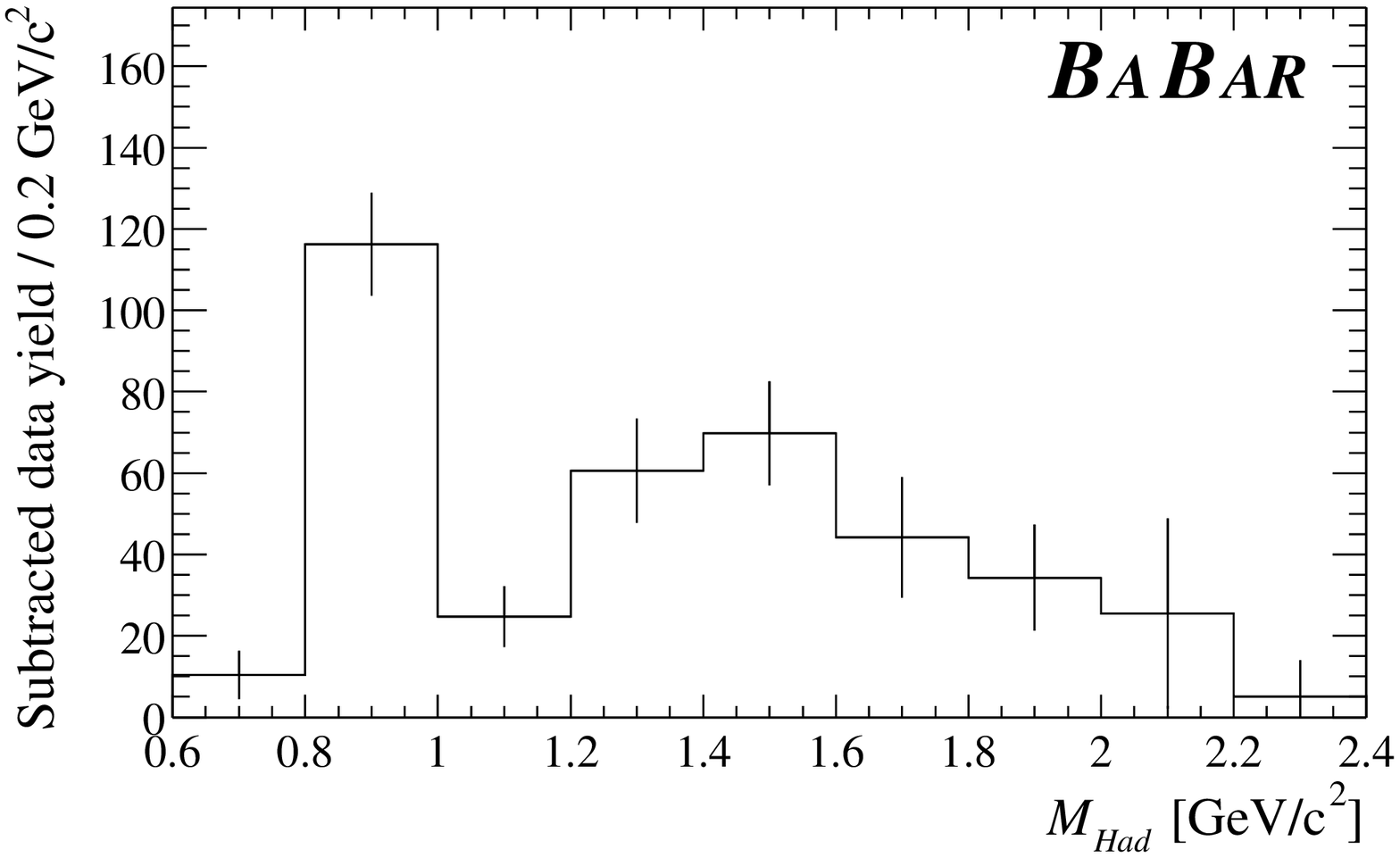}} &
   \mbox{\includegraphics[width=3.2in]{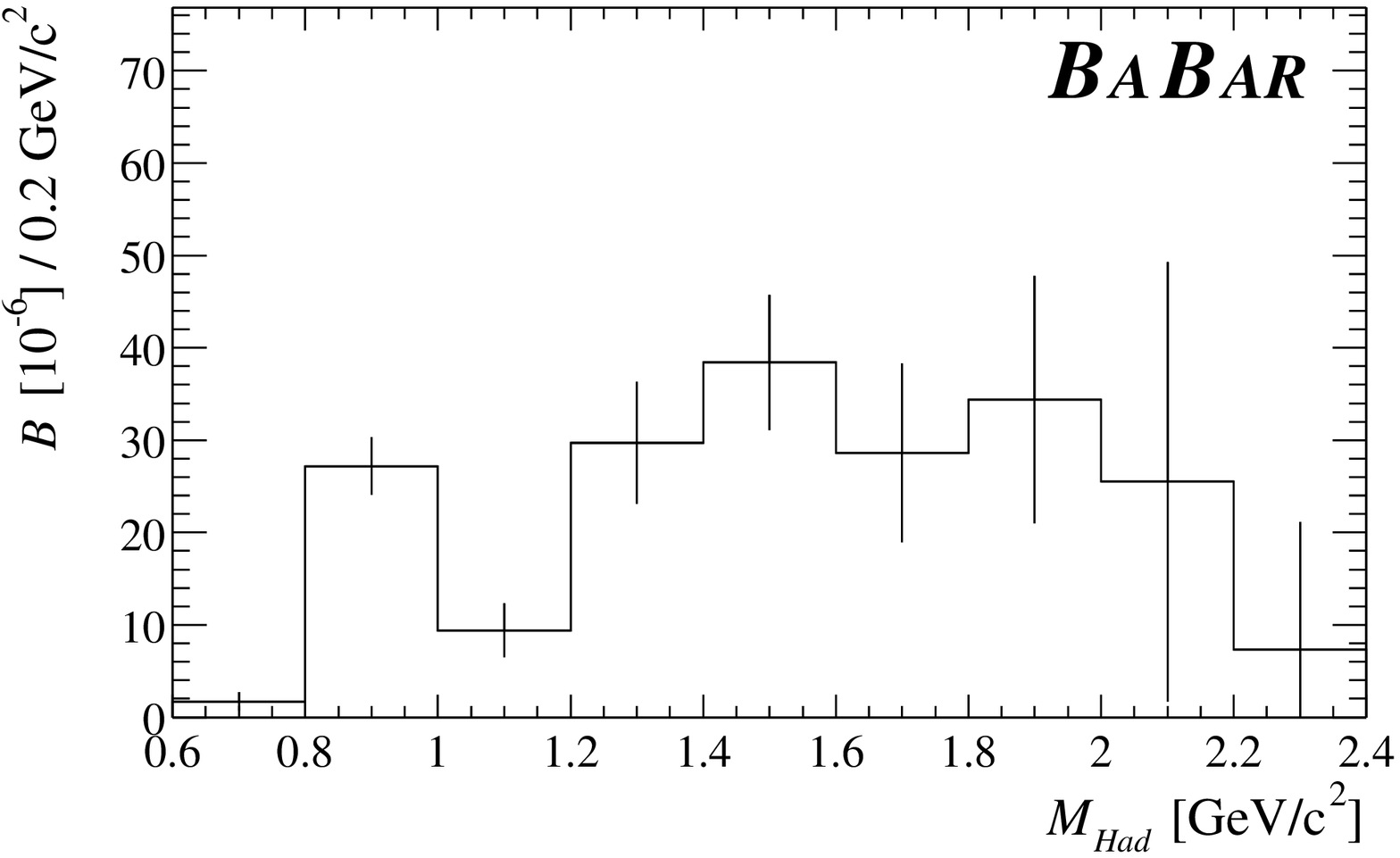}} \\
  \end{tabular}
\caption{Signal yields (left plot) and weighted partial branching fractions (right plot), calculated with 
corrected efficiencies. Statistical errors only are shown.
\label{fig:pbf_all}}
\end{figure}

Partial branching fractions ({$\Delta \cal B$}) are calculated for each \mhad\ bin
for the sum of the twelve final states reconstructed in this analysis.
The calculation uses the data and \BB\ yields from Table~\ref{tab:allyield} and the
signal efficiencies from Table~\ref{tab:effsig}. The peaking cross--feed background has to be
normalized to our measured inclusive branching fraction (${\cal B}$).
We take care of this dependence of the peaking cross--feed 
background by including effective terms that modify the signal efficiencies in each bin,
and report the equivalent peaking cross--feed yields in Table~\ref{tab:allyield}.
Figure~\ref{fig:pbf_all} shows the data yields after subtraction of the peaking  
\BB\ and cross--feed backgrounds (signal yields). 

The inclusive branching fraction for $b\to s\gamma$ in each \mhad\ bin
can be obtained from the partial branching fractions by taking into account 
the fraction, $f$, of final states that are considered in the analysis.
The $f$s are taken from the generator level JETSET fragmentation of the 
inclusive $b\to s\gamma$ signal.  
Since there are discrepancies in the modeling of our 
twelve reconstructed final states as already noted above, we know that the values of $f$ 
are not completely reliable. 
There are also uncertainties coming from other final states,  
e.g. with $\eta$ and $\phi$ mesons, which have not been looked for 
and therefore cannot be compared with data.

We divide up the missing final states from our generator--level Monte Carlo
into categories with a $K_L$ (25\%), more than one $\pi^0$ (4-18\%), 
more than 4--body hadronic systems (0-14\%),
an $\eta/\eta'$ (5-11\%), a $s\bar{s}s$ system (0-7\%), and ``other''(0-4\%), where 
the missing fractions increase from $\mhad=1.0$ to $\mhad=2.4$~\gevcc. 
The $K_L$ fraction is known to be 25\% from isospin and the measured 
$K_S/K^+$ ratio in data. For all the other fractions we allow a variation 
between 0.5 and 1.5 times the generator fraction, and add these variations 
in quadrature to give an overall systematic uncertainty on the missing fractions.
Table~\ref{tab:fi} summarizes the partial branching fractions with their statistical errors, 
the fraction of studied final states with the corresponding errors, the resulting branching fractions
with their statistical and systematic errors and the cumulative $b\to s\gamma$ branching fraction,
with statistical and systematic errors, in bins of \mhad. 

\begin{table}[htbp] 
\begin{center}
\begin{tabular}{|c|rcl|rcl|rcccl|rcccl|} \hline
\mhad~[\gevcc]& \multicolumn{3}{|c|}{{$\Delta \cal B$} $\cdot 10^{-6}$} 
& \multicolumn{3}{|c|}{$f$ (\%)}& \multicolumn{5}{|c|}{${\cal B}$ $\cdot 10^{-6}$}  
& \multicolumn{5}{|c|}{Cumulative ${\cal B}$ $\cdot 10^{-6}$}  \\ \hline
0.6-0.8    &   2 & $\pm$ &  1   &    &   75  &    & 2.3 & $\pm$ & 1.3 & $\pm$ & 0.5 &   2.3 & $\pm$ &   1.3 & $\pm$ &  0.5  \\
0.8-1.0    &  27 & $\pm$ &  3   &    &   75  &    &  36 & $\pm$ &   4 & $\pm$ &  5  &  39   & $\pm$ &   4   & $\pm$ &  5       \\
1.0-1.2    &   9 & $\pm$ &  3   & 66 & $\pm$ & 3  &  14 & $\pm$ &   4 & $\pm$ &  4  &  53   & $\pm$ &   6   & $\pm$ &  7    \\
1.2-1.4    &  30 & $\pm$ &  7   & 63 & $\pm$ & 3  &  47 & $\pm$ &  10 & $\pm$ &  9  & 100   & $\pm$ &  12   & $\pm$ & 14     \\
1.4-1.6    &  38 & $\pm$ &  7   & 55 & $\pm$ & 4  &  70 & $\pm$ &  13 & $\pm$ & 13  & 170   & $\pm$ &  18   & $\pm$ & 25      \\
1.6-1.8    &  29 & $\pm$ & 10   & 45 & $\pm$ & 7  &  64 & $\pm$ &  22 & $\pm$ & 15  & 233   & $\pm$ &  28   & $\pm$ & 36      \\
1.8-2.0    &  34 & $\pm$ & 13   & 35 & $\pm$ & 8  &  98 & $\pm$ &  38 & $\pm$ & 30  & 332   & $\pm$ &  48   & $\pm$ & 61    \\
2.0-2.2    &  26 & $\pm$ & 24   & 27 & $\pm$ & 10 &  95 & $\pm$ &  88 & $\pm$ & 43  & 426   & $\pm$ & 100   & $\pm$ & 98     \\
2.2-2.4    &   7 & $\pm$ & 14   & 22 & $\pm$ & 11 &  33 & $\pm$ &  63 & $\pm$ & 62  & 460   & $\pm$ & 118   & $\pm$ & 134     \\
\hline 
\end{tabular}
\end{center}
\caption{The partial branching fractions ({$\Delta \cal B$}), the fraction of final states that 
are considered in the analysis ($f$), the total $b\to s\gamma$ 
branching fractions ({$\cal B$}) and the cumulative total $b\to s\gamma$ ${\cal B}$ in bins of \mhad. 
Statistical errors are shown for the {$\Delta \cal B$}s, and 
systematic uncertainties for the $f$s. Both statistical and overall systematic errors
are shown for the {$\cal B$}s.}
\label{tab:fi}
\end{table}

The systematic errors will be described in Section~\ref{sec:systerr}.
It can be seen that the branching fraction for $\mhad < 1.0$~\gevcc\ agrees
with the exclusive $K^*\gamma$ branching fraction of about $4\cdot 10^{-5}$~\cite{Kstar},
and that the integral for $\mhad > 2.0$~\gevcc\ is higher than previous measurements
of the inclusive $b\to s\gamma$ branching fraction~\cite{CLEO, BELLE}, 
although with large uncertainties. Note that these results are essentially 
independent of the modeling of the $b\to s\gamma$ spectrum. 

\begin{figure}[htb]
\begin{center}
   \mbox{\includegraphics[width=4.2in]{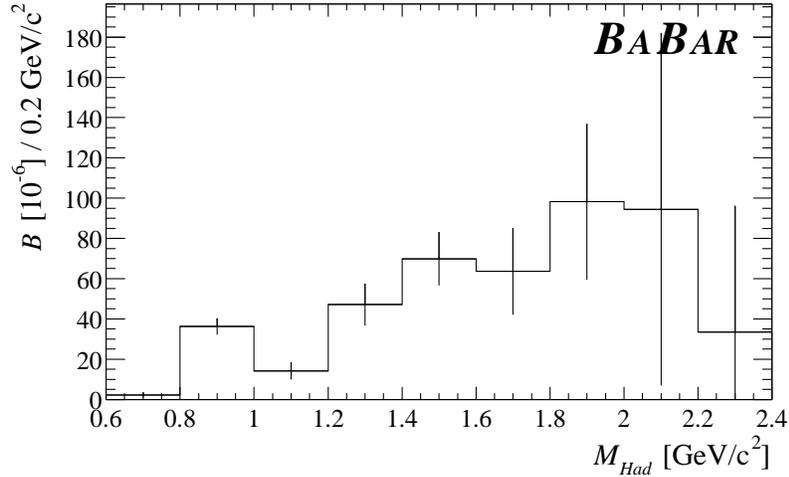}} \\
\end{center}
\vspace{-0.8cm}
\caption{Branching fraction as a function of \mhad. 
The errors are purely statistical.
\label{fig:inclusive}}
\end{figure}

\begin{figure}[hbt]
\begin{center}
   \mbox{\includegraphics[width=4.2in]{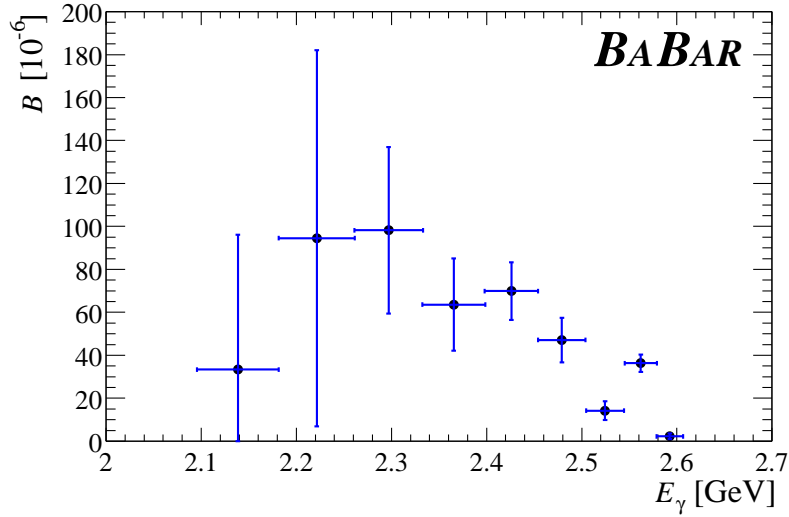}} \\
\end{center}
\vspace{-1.0cm}
\caption{Branching fraction as a function of $E_{\gamma}$.
The errors are purely statistical.
\label{fig:inclusiveE}}
\end{figure}

Figure~\ref{fig:inclusive} shows the branching fractions as a function of \mhad. The same results 
are shown as a function of $E_{\gamma}$ in Figure~\ref{fig:inclusiveE}. 

\section{Systematic Errors}
\label{sec:systerr}

The first category of systematic errors ({\it general systematic}) are those which do not depend on the hadronic final state.
The systematic error on the detection of the high--energy photon 
is 3.7\%, and the systematic error on the total number of \B\ 
mesons in our data sample is 1.6\%. 
There is also a systematic uncertainty of 1\%\ coming from the use of 
the $\cost$ and Fisher Discriminant requirements to suppress continuum backgrounds.
These errors give a total contribution of 4.2\%\ independent of the hadronic mass bin.
 
A second category of errors are due to the fitting procedure and signal 
definition ({\it fit systematic}). 
These errors are evaluated by varying the 
fixed parameters used in the fits, and by varying the 
fit procedures. The width of the signal Crystal Ball function is varied by 
$\pm$0.2~\mevcc\ for both signal and peaking backgrounds. 
This introduces a systematic uncertainty of 5\% on the signal yield. 
Varying the Crystal Ball shape to reflect our understanding of the radiative tail
introduces a further systematic uncertainty of 5\%, and varying 
the peak position introduces a systematic uncertainty of 3\%.
The ARGUS shape parameters are varied within the ranges allowed by the 
Monte Carlo samples. They give a contribution to the systematic 
uncertainty of 10\%, which is limited by the Monte Carlo statistics.
We also include a systematic error of 20\% on the peaking \BB\ yield
in the region \mhad = 1.8-2.4~\gevcc\ due to the method of fitting 
the \BB\ yield, and to uncertainties in the content of the generic \BB\ \MC\
simulation. 
This translates into a 10\% systematic error in the last three \mhad\ bins. 

Next we consider errors in the detection efficiencies for the hadronic final states ({\it efficiency systematic}).
We assign systematic errors based on the uncertainties in the corrections to the efficiencies 
that we have applied. The uncertainties coming from the single particle detection differences 
between data and \MC\ are between 4.0 and 4.6\% as a function of \mhad, 
and the fragmentation differences give uncertainties between 2.1 and 6.3\%.

A final category of errors is due to the generator and modeling we use ({\it generator systematic}).
The errors due to the missing final states not considered in the analysis
were discussed in the previous section and shown in Table~\ref{tab:fi}. 
Also, we consider contributions due to changing the modeling of the bin 1.0-1.2~\gevcc\ to $K^*\gamma$ and varying
the parameters of the inclusive model in our signal Monte Carlo. 

The contributions to the systematic errors coming from all sources are summarized 
in Table~\ref{tab:summarySyst}. 

\begin{table}[htbp]
\begin{center} 
\begin{tabular}{|c|c|c|c|c|c|c|c|c|}\hline     
\mhad  &{\it general           }&{\it  fit          }&{\it  efficiency   }&{\it   generator} & total    \\  
$[$\gevcc$]$   &{\it  systematic} (\%)&{\it   systematic} (\%) &{\it   systematic} (\%) &{\it   systematic} (\%) & systematic (\%) \\ \hline
0.6-0.8 &  4.2 &  17 &  4.0 & --  &  18    \\
0.8-1.0 &  4.2 &   6 &  4.0 & --  &   8     \\
1.0-1.2 &  4.2 &  24 &  5.1 & 15  &  29     \\
1.2-1.4 &  4.2 &  12 &  7.7 & 11  &  18 \\
1.4-1.6 &  4.2 &  13 &  5.5 & 12  &  19     \\
1.6-1.8 &  4.2 &  12 &  7.2 & 19  &  24     \\
1.8-2.0 &  4.2 &  17 &  5.1 & 25  &  31     \\
2.0-2.2 &  4.2 &  23 &  6.4 & 38  &  45    \\
2.2-2.4 &  4.2 & 177 &  5.1 & 51  &  184   \\
\hline
\end{tabular}
\end{center}
\caption{All contributions to the systematic error are shown together with the total sum in quadrature as a function of the hadronic mass interval.
\label{tab:summarySyst}}
\end{table}

\section{Fits to the Spectrum}

So far our results have had no significant dependence on the $b\to s\gamma$ model.
We now fit the hadronic mass spectrum with the shape predicted 
by the Kagan and Neubert model~\cite{KaganNeubert}. This extrapolates 
from the measured range of \mhad\ to give the inclusive branching fraction 
for $b\to s\gamma$.

\begin{figure}[htb]
\begin{center} 
   \mbox{\includegraphics[width=5.in]{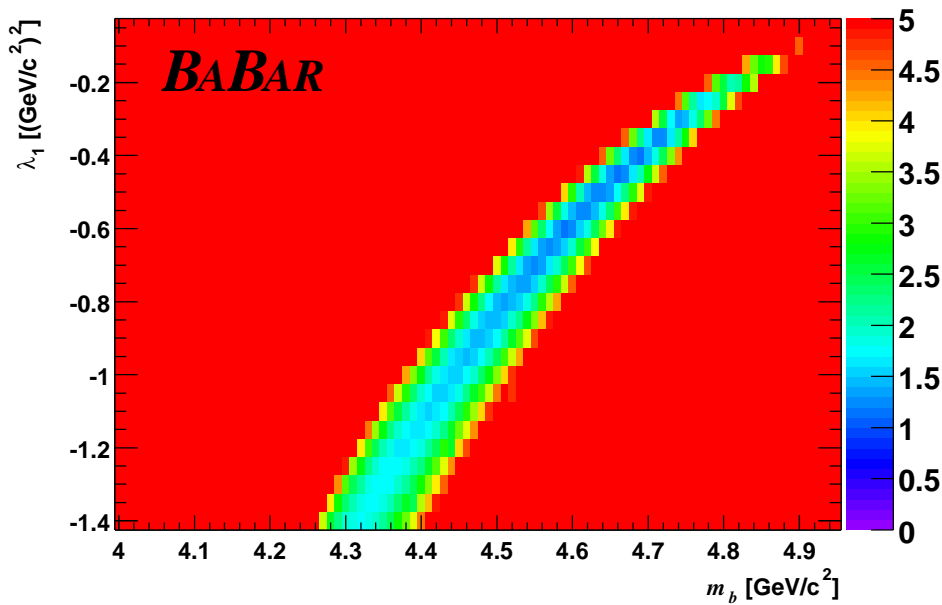}} \\ 
\end{center}
\vspace{-1.0cm}
\caption{Minimum $\chi^2$ value in the plane of $\lambda_1, m_b$. 
Values of the $\chi^2$ larger than 5 are set to 5.
Note the very strong correlation between $\lambda_1$ and $m_b$. 
\label{fig:color}}
\end{figure}

\begin{figure}[htbp]
\begin{center}
\begin{tabular}{c} 
   \mbox{\includegraphics[width=3.5in]{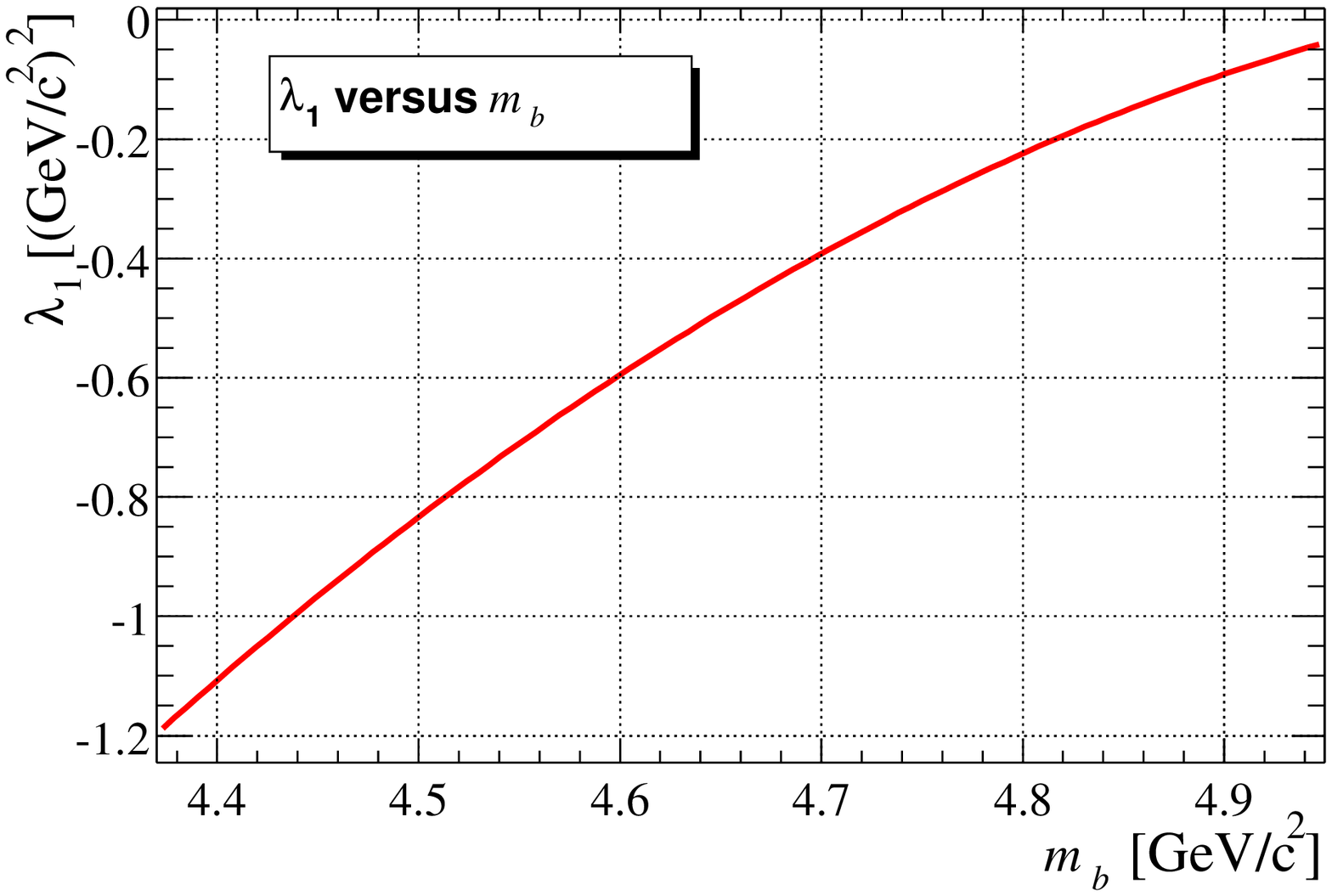}} \\
   \mbox{\includegraphics[width=3.5in]{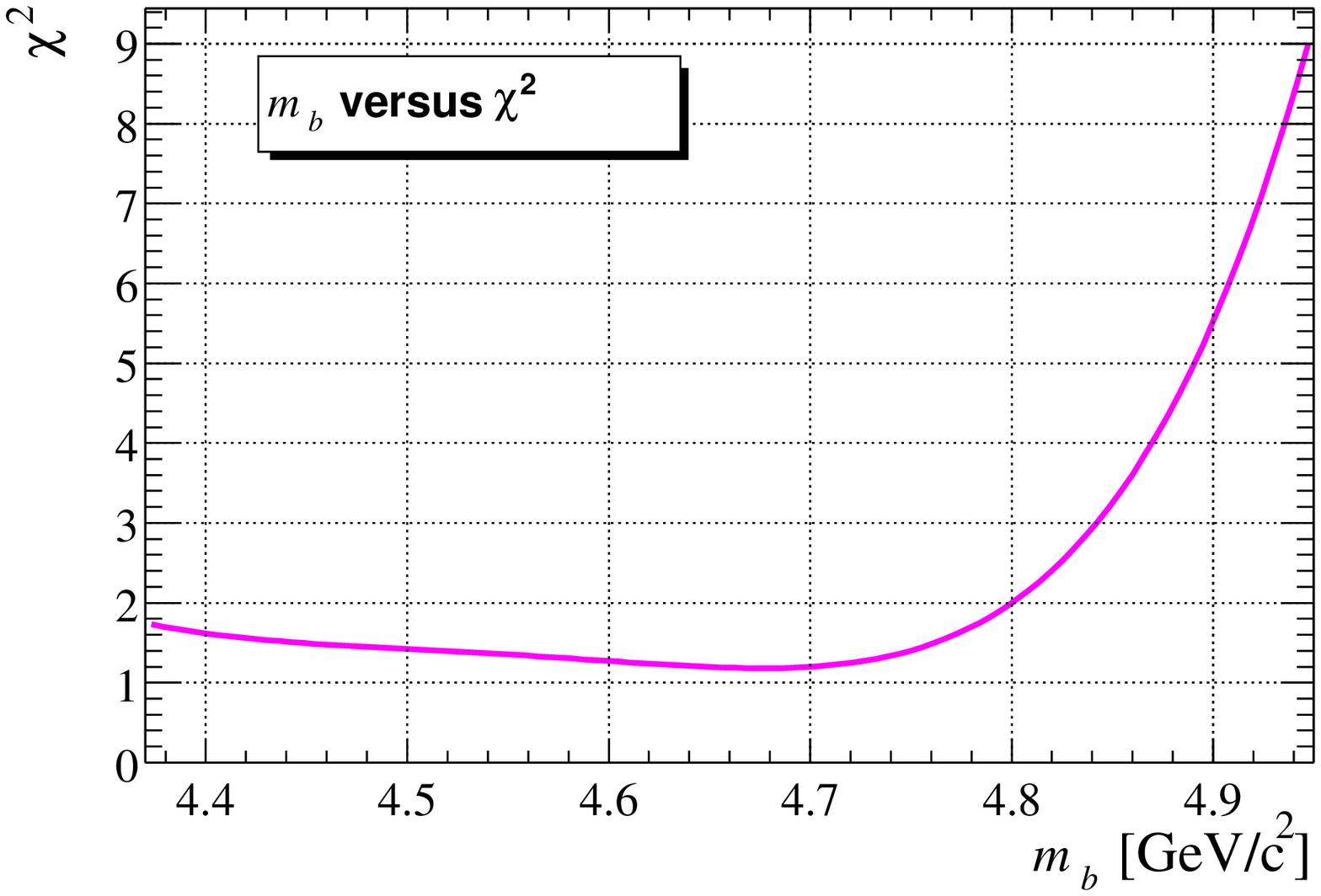}} \\
   \mbox{\includegraphics[width=3.5in]{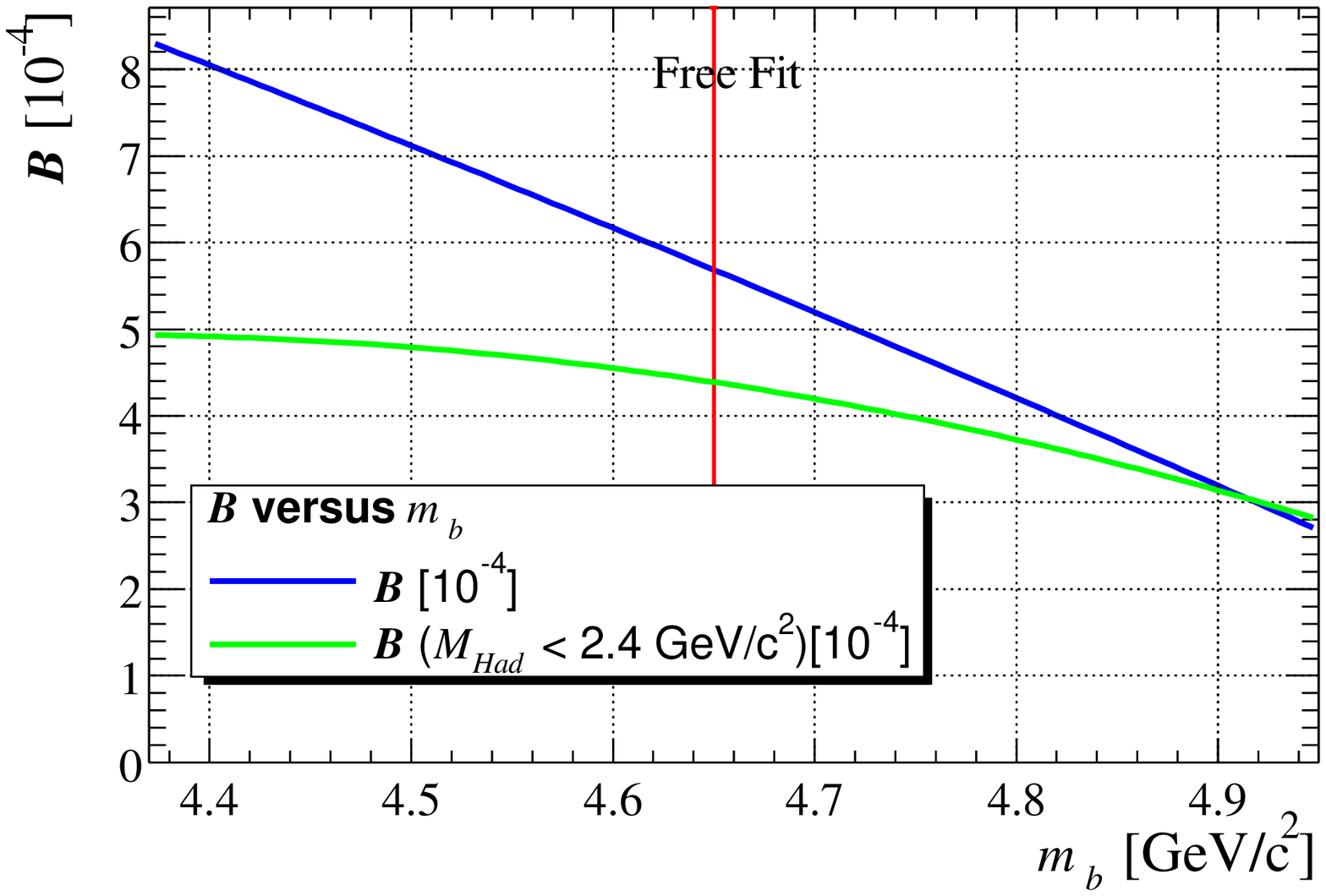}} \\
\end{tabular}
\end{center}
\vspace{-0.8cm}
\caption{Results from a free fit to the \mhad\ spectrum.
Plot (a) shows the correlation curve between  $\lambda_1$ and $m_b$ along which 
the $\chi^2$ is minimized. 
Plot (b) shows the minimum $\chi^2$ as a function of $m_b$.  
Plot (c) shows the inclusive ${\cal B}$ and the ${\cal B}$ for \mhad\ $<$ 2.4~\gevcc\ as a function of $m_b$.  
\label{fig:chi2min}}
\end{figure}

In the fits we vary the parameters $m_b$ and $\lambda_1$,
the transition point between the $K^*$ and non-resonant contributions, 
and the total normalization, which corresponds to our integrated branching fraction. 
In the plane $\lambda_1$ and $m_b$, Figure~\ref{fig:color} shows 
the value of the $\chi^2$, having fit the transition point (resulting around 
1.1~\gevcc) and the normalization for each point.
There is a strong correlation between these parameters, which is expected in this model, 
and a shallow minimum at $m_b=4.65$~\gevcc\ and $\lambda_1=-0.48 \rm ~[\gevcc]^2$.
Figure~\ref{fig:chi2min}(a) shows the correlation curve between $\lambda_1$ and 
$m_b$ along which we have the minimum value of $\chi^2$, and 
Figure~\ref{fig:chi2min}(b) shows the values of $\chi^2$ as a 
function of $m_b$. Figure~\ref{fig:chi2min}(c) shows the inclusive branching fraction 
and the branching fraction for \mhad\ $<$ 2.4~\gevcc\ 
as a function of $m_b$, where the result for the minimum value of $m_b$ found in the free 
fit is marked with a line. Since our spectrum contains a $K^*$ peak with a known branching 
fraction of about $4\cdot 10^{-5}$, a reduction in $m_b$ leads to an 
increase in the non-resonant
component of the spectrum, and hence to a higher inclusive branching fraction. 
This increase can be seen both in the branching fraction restricted 
to the measured range, and in the extrapolation over all \mhad.


The free fit to the model parameters only allows us to set an upper bound on $m_b$, since 
the $\chi^2$ rises rather slowly on the other side of the minimum.
We obtain a lower bound on $m_b$ by measuring the first moment of the 
$E_\gamma$ distribution, $<E_\gamma>$, and then extracting 
${\rm \overline{\Lambda}}(\alpha^2_s,1/M_B^3)$ 
using HQET calculations~\cite{ligeti,bauer} with the coefficients relevant
for our minimum value of $E_\gamma=2.094$~\gev~\cite{ligetiemail}.
Then, we truncate the HQET expressions to order $O(\alpha_s,1/M_B^2)$ and recompute 
$m_b = M_B - {\rm \overline{\Lambda}}(\alpha_s,1/M_B^2)$ to obtain 
the parameter required in the 
Kagan and Neubert model. There are significant theoretical uncertainties associated 
with this method coming from varying the scale at which $\alpha_s$ is determined between 
$m_b/2$ and $2m_b$, from the truncation of higher--order terms, 
and from the dependence of $\rm \bar{\Lambda}$ on the choice of the minimum $E_{\gamma}$~\cite{bauer}. 

The value of the first moment, $<E_\gamma>|_{\tiny E_{\gamma}>2.094~\gev}$, is: 
\begin{center}
$<E_\gamma>|_{\tiny E_{\gamma}>2.094~\gev} = 2.35 \pm 0.04~(stat) \pm 0.04~(syst)$~\gev, 
\end{center}
which gives: 
\begin{center}
${\rm \overline{\Lambda}} = 0.37 \pm 0.09~(stat) \pm 0.07~(syst) \pm 0.10~(model)$~\gevcc. 
\end{center}
where the theoretical uncertainties have been included $(model)$.
With the truncation to first order in $\alpha_s$, ${\rm \overline{\Lambda}} = 0.49$~\gevcc, which gives: 
\begin{center}
$m_b=4.79 \pm 0.08~(stat) \pm 0.10~(syst) \pm 0.10 ~(model)$~\gevcc.
\end{center}

\section{Final Results}

We perform final fits to the hadronic mass spectrum with 
$m_b= 4.79\pm 0.16\gevcc$ taken from the first moment analysis.
These fits give the inclusive branching fraction:
\begin{center}
${\cal B} (b\to s\gamma)        = 4.3  \pm  0.5 ~(stat)   \pm 0.8 ~(syst) \pm 1.3 ~(model) \cdot 10^{-4}$ 
\end{center}
and a range for the parameter $\lambda_1$:
\begin{center}
$\lambda_1    = -0.24 ^{+0.03} _{-0.04}~(stat) \pm 0.02 ~(syst) ^{+0.15} _{-0.21} ~(model)~[\gevcc]^2$.
\end{center}

\begin{figure}[hb]
\begin{center} 
   \mbox{\includegraphics[width=5.in]{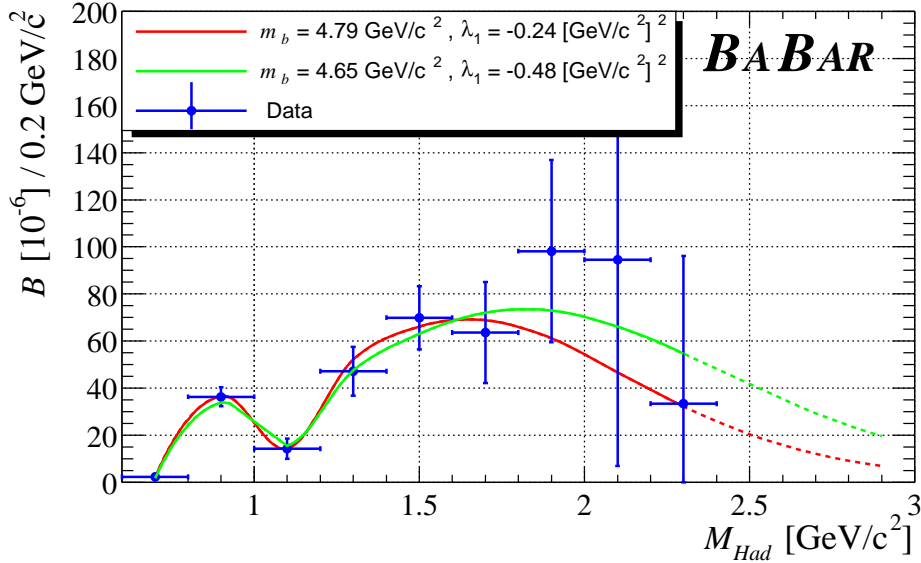}} 
\end{center}
\vspace{-0.8cm}
\caption{Superposition of the predicted spectrum for 
$m_b  = 4.79$~\gevcc\ and $m_b=4.65$~\gevcc\ on the observed hadronic mass spectrum.
\label{fig:onedim}}
\end{figure}

Figure~\ref{fig:onedim} shows the superposition of the theoretical spectrum on the data distribution 
for the best fits with $m_b=4.79$~\gevcc\ and $m_b=4.65$~\gevcc. 

Our values for ${\rm \overline{\Lambda}}$, at $O(\alpha^2_s,1/M_B^3)$, and $\lambda_1$, 
at $O(\alpha_s,1/M_B^2)$, are consistent with other results~\cite{CLEOmoment}, 
but our inclusive branching fraction is somewhat higher~\cite{CLEO,BELLE}. 
Our error bars are still larger than the previous measurements, 
but they can be reduced significantly with the addition of more data.
It is possible to reduce the systematic errors by improvements 
in the treatment of backgrounds, and by studying a larger number 
of final states to reduce the missing fraction errors.  
The model dependence will also be reduced by fitting the spectrum to 
determine the model parameters. 

\section{Acknowledgments}
We would like to thank Alexander L. Kagan, Zoltan Ligeti and Mark B. Wise, for useful discussions and for 
having provided us with numbers and results from their works. 
We are grateful for the 
extraordinary contributions of our \pep2\ colleagues in
achieving the excellent luminosity and machine conditions
that have made this work possible.
The success of this project also relies critically on the 
expertise and dedication of the computing organizations that 
support \babar.
The collaborating institutions wish to thank 
SLAC for its support and the kind hospitality extended to them. 
This work is supported by the
US Department of Energy
and National Science Foundation, the
Natural Sciences and Engineering Research Council (Canada),
Institute of High Energy Physics (China), the
Commissariat \`a l'Energie Atomique and
Institut National de Physique Nucl\'eaire et de Physique des Particules
(France), the
Bundesministerium f\"ur Bildung und Forschung and
Deutsche Forschungsgemeinschaft
(Germany), the
Istituto Nazionale di Fisica Nucleare (Italy),
the Research Council of Norway, the
Ministry of Science and Technology of the Russian Federation, and the
Particle Physics and Astronomy Research Council (United Kingdom). 
Individuals have received support from 
the A. P. Sloan Foundation, 
the Research Corporation,
and the Alexander von Humboldt Foundation.

\end{document}